\documentclass[a4paper,11pt,twoside]{article}

\usepackage{amsmath}
\usepackage{amssymb}
\usepackage{epsfig}
\usepackage{wrapfig}
\usepackage{array}

\newcommand{\ass}{\renewcommand{\arraystretch}{1}}
\newcommand{\asl}{\renewcommand{\arraystretch}{1.2}}

\renewcommand{\t}[1]{\theta_{#1}}
\newcommand{\mee}{m_{ee}}
\newcommand{\bb}{0\nu 2\beta}
\newcommand{\U}{V_\nu}
\newcommand{\mn}{M_N}
\newcommand{\mnf}{M_\nu}
\renewcommand{\L}{{\cal L}^\nu_m}
\newcommand{\hc}{\text{h.c.}}

\newcommand{\ord}[1]{\mathcal{O}\left( #1 \right)}

\newcommand{\dm}[1]{{\Delta m^2_{#1}}}
\newcommand{\Fig}[1]{Fig.~\ref{fig:#1}}

\newcommand{\eq}[1]{eq.~(\ref{eq:#1})}

\newcommand{\capdef}{}
\newcommand{\mycaption}[2][\capdef]{\renewcommand{\capdef}{#2}%
        \caption[#1]{{\itshape #2}}} 
\makeatletter
\renewcommand{\fnum@table}{\textbf{\tablename~\thetable}}
\renewcommand{\fnum@figure}{\textbf{\figurename~\thefigure}}
\makeatother

\newlength{\myem}
\newcounter{mysubequation}[equation]

\makeatletter
\renewcommand{\section}{\@startsection{section}{1}{0em}%
        {-3.5ex \@plus -1ex \@minus -.2ex}%
        {2.3ex \@plus.2ex}%
        {\normalfont\large\bfseries}}
\renewcommand{\subsection}{\@startsection{subsection}{2}{0em}%
        {-3.25ex\@plus -1ex \@minus -.2ex}%
        {1.5ex \@plus .2ex}%
        {\normalfont\bfseries}}
\renewcommand{\subsubsection}%
        {\@startsection{subsubsection}{3}{0em}%
        {-3.25ex\@plus -1ex \@minus -.2ex}%
        {1.5ex \@plus .2ex}%
        {\normalfont\bfseries}}
\makeatother

\newcommand{\CERN}{Theory Division, CERN, CH--1211 Geneva, Switzerland}

\newcommand{\SNS}{Scuola Normale Superiore and INFN Sezione di Pisa \\ 
Piazza dei Cavalieri 7, I--56126 Pisa, Italy}

\newcommand{\preprintdate}{}
\newcommand{\preprintnumber}{CERN-TH/2003-027}
 
\newcommand{\titletext}{Natural relations among physical observables
  in the neutrino mass matrix} 
\newcommand{\authortext}{\large Riccardo Barbieri$^{\, a}$, Thomas
  Hambye$^{\, a}$ and Andrea Romanino$^{\, b}$ 
\medskip\\\em\normalsize 
$\mbox{}^a$ \SNS
\\[0.1\baselineskip] 
$\mbox{}^b$ \CERN}
\newcommand{\abstracttext}{We find all possible relations
  among physical observables arising from neutrino mass matrices that
  describe in a natural way the currently observed pattern ($\t{23}$ and
  $\t{12}$ large, $\dm{\odot}/\dm{\text{Atm}}$ and $\t{13}$ small) in terms 
  of a minimum number of parameters.
  Natural here means due only to the relative smallness (vanishing) of
  some parameters in the relevant lagrangian, without special
  relations or accidental cancellations among them.}

\title{
\normalsize
\begin{tabular}[t]{l}
\preprintdate\end{tabular}
\hspace*{\fill}
\begin{tabular}[t]{l}\preprintnumber\end{tabular}
\vspace{3\baselineskip}\\\Large\bfseries\titletext\bigskip}
\author{\begin{minipage}[t]{0.8\textwidth}
\normalsize\centering\authortext
\end{minipage}}
\date{}

\begin{document}

\bigskip
\maketitle
\begin{abstract}\normalsize\noindent
\abstracttext
\end{abstract}\normalsize\vspace{\baselineskip}


\section{Introduction}

To date, 3 different kinds of experiments are being used or can be
conceived to get informations on the physical parameters in the mass
matrix of 3 Majorana neutrinos: i) oscillation experiments, which can
make access to the 2 independent squared mass differences $\dm{32}$,
$\dm{21}$ ($|\dm{32}|>\dm{21}>0$), to the 3 mixing angles $\t{23}$,
$\t{12}$, $\t{13}$, and to a CP violating phase $\delta$; ii) $\beta$
and $\bb$ decay experiments, probing 2 appropriate combinations of
masses and oscillation parameters, $m_\beta$ and $\mee$; iii)
cosmological and/or astrophysical experiments, which might detect,
e.g., the sum of the neutrino masses $\sum_i m_i$.  An increasing
level of difficulty is involved in the different experiments, however,
to reach the needed sensitivity, crucially depending on the actual
neutrino mass spectrum.  For a hierarchical spectrum, either
``normal'' ($m_3\gg m_2> m_1$) or ``inverted'' ($m_1\simeq m_2\gg
m_3$), as we shall consider in the following\footnote{The degenerate
  spectrum is not considered since it does not satisfy the naturalness
  criterion as defined below.}, one can tentatively assume, at least
for illustration purposes, that all the six oscillation observables
will be measured, some of them with significant precision, perhaps
together with the $\bb$ mass $\mee$.  This makes a total of seven
observables, which a theory of neutrino masses should be able to
correlate among each other.

A simple minded conjecture is that some of these correlations could
arise from the economy in the number of independent basic parameters.
As the simplest example, one may consider the possibility that the
neutrino mass matrix $\mnf$, in the flavour basis, has a maximum
number of negligibly small entries~\cite{Frampton:02a,Guo:02a}.
Inspection shows that, out of the 6 independent elements of the
$3\times 3$ symmetric matrix $\mnf$, only 2 of them could at most
vanish consistently with current
observations~\cite{kamland:02a,Apollonio:02a}
\begin{equation}
  \label{eq:obs}
  \begin{split}
  0.02 < R\equiv \dm{21}/\dm{32} &< 0.04, \quad 0.35 < \tan^2\t{12} <
  0.55, \\ 
  0.6 < \tan^2\t{23} < 1.4, &\quad
  \sin^2\t{13}\lesssim 0.03. 
\end{split} 
\end{equation}
This leaves a total of 15 different mass matrices to be examined, all
of which dependent on 4 real parameters and 1 phase that can affect
the 6 oscillation observables and $\mee$. One expects therefore 2
relations between these observables, a priori different in every case.

Although interesting, the limit of this approach is that the elements
of the neutrino mass matrix in the flavor basis are, in general, only
combinations of the basic parameters in the relevant piece of the
lagrangian, $\L$. For sure $\mnf$ is generally influenced also by the
charged lepton mass matrix or, if the seesaw mechanism is operative,
by the mass matrix of the right handed neutrinos. A natural question
to ask therefore is what happens if the economy in the number of
parameters is not in $\mnf$ but rather in the basic lagrangian. This
is the question we address in this paper.

To answer this question in full generality would require examining a
huge number of different possibilities, of which the 15 cases for
$\mnf$ mentioned above are only a small subset. In the following we
shall only consider the possibilities that describe the currently
observed pattern of the data~(\ref{eq:obs}) in a ``natural'' way,
i.e.\ only by the relative smallness (vanishing) of some parameters in
$\L$, barring special relations or accidental cancellations among
them. This should in particular be the case when accounting
simultaneously for the smallness of $R\equiv \dm{21}/\dm{32}$ and for
the largeness of $\t{23}$, the most peculiar feature of the data so
far, even though an accidental cancellation might also produce the
same feature.  As illustrated below, it turns out that none of the 15
cases for $\mnf$ mentioned above satisfy this criterion in a strict
sense. Note that naturality is not defined here in terms of any
symmetry. In particular we do not require that the relative smallness
(vanishing) of some parameters in the basic lagrangian be understood
in terms of an explicit symmetry, exact or approximate. The coming
into play of symmetries could add new cases to our list. Viceversa, it
could prove hard to understand some of the cases we consider as due to
a flavour symmetry of any sort.

As anticipated, we shall independently consider:
\smallskip \\
i) the direct non see-saw case (NSSC) for the 3 Majorana neutrinos
$N^T_L = (n_1,n_2,n_3)$, described by the mass lagrangian (after
electroweak symmetry breaking)
\begin{equation}
  \label{eq:LNSSC}
  \L(\text{NSSC}) = \bar{E_L} M_E E_R+N^T_L \mn N_L +\hc,
\end{equation}
where $E_L^T=(e,\mu,\tau)_L$, $E_R^T = (e,\mu,\tau)_R$;
\smallskip \\
ii) the see-saw case (SSC) with the lagrangian mediated by 2 or 3
right handed neutrinos\footnote{In a 3 neutrino context, the 2
  neutrino case corresponds to the limit in which one or more entries
  of the heavy neutrino mass matrix become much larger than the
  others.}, collectively denoted by $N$
\begin{equation}
  \label{eq:LSSC}
  \L(\text{SSC}) = \bar{E_L} M_E E_R + N^T M_{RL} N_L + N^T M_{R}
  N +\hc.
\end{equation}

The neutrino mass matrix in the flavour basis, $\mnf$, and the
leptonic analog of the Cabibbo Kobayashi Maskawa matrix $\U$ are given
respectively by 
\begin{equation}
  \label{eq:rotations}
  \mnf = U^T_l \mn U_l,\qquad \U = U^\dagger_l U_\nu
\end{equation}
in terms of the diagonal matrices 
\begin{equation}
  \label{eq:diag}
  D_E = U^\dagger_l M_E V_l, \qquad D_N = U_\nu^T \mn U_\nu,
\end{equation}
where, in the SSC, $-\mn = M_{RL}^T M^{-1}_R M_{RL}$. To count easily
the number of effective parameters in $M_N$, in the SSC it is
convenient to define
\begin{equation}
  \label{eq:A}
  M_{RL} = m A
\end{equation}
where $m_{ij} = m_i\delta_{ij}$ and $(AA^\dagger)_{ii} = 1$ with
$i=1,2$ ($1,2,3$) for 2 (3) right handed neutrinos. Note that this
decomposition is unique, with all $m_i$'s positive. It is then
\begin{equation}
  \label{eq:mu}
  -\mn = A^T\mu^{-1} A \quad\text{with}\quad \mu^{-1} \equiv m M^{-1}_R m
\end{equation}
($A$ is adimensional, while $\mu$ has the dimension of an inverse
mass). Of course the naturalness requirement applies equivalently to
$M_{R}$, $M_{RL}$ or to $\mu$, $A$.  The number of effective
parameters in $M_N$ is the sum of the parameters in $\mu$ and $A$.

Within this framework, in the next section we describe all cases that
i) are consistent with current data, as summarized in~(\ref{eq:obs});
ii) lead to some definite testable correlation between the different
observables.\footnote{A case which leads to $\sin\t{13}=0$, $m_{ee}=0$
  is disregarded, since we do not see how it can receive experimental
  support.} More precisely, to realize ii) we stick to the cases that
give a neutrino mass matrix in the flavor basis involving at most 4
real parameters and one phase.

\section{Summary of results}
\label{sec:cases}

\asl
\begin{table}[t]
  \centering
  \[\begin{array}{|c||c|c|c|c|c|c|} 
    \hline
    &
    \sin\t{13} &
    |m_{ee}|/m_{\text{atm}} &
    \text{SSC} &
    \hspace*{-1mm}\text{NSSC}\hspace*{-1mm} &
    U_l \\
    \hline\hline
    \text{A1}&
    \displaystyle\rule[-0.4cm]{0cm}{1.1cm}
    \frac{1}{2}\tan\t{23}\sin2\t{12}\,\sqrt{R} &
    \sin^2\t{12}\sqrt{R} &
    \text{2} &
    &
    \mathbf{1} \\
    \hline
    \text{B1} &
    \displaystyle\rule[-0.4cm]{0cm}{1.1cm}
    \frac{1}{2}\tan\t{23}\tan2\t{12}\,(R\cos2\t{12})^{1/2} &
    0 &
    \text{3} &
    &
    \mathbf{1} \\
    \hline
    \text{C} &
    \displaystyle\rule[-0.4cm]{0cm}{1.1cm}
    \frac{1}{2}\tan2\t{12}\,(R\cos2\t{12})^{3/4} &
    0 &
    \text{3} &
    \surd &
    R_{23} \\
    \hline
    \text{D} &
    \displaystyle\rule[-0.4cm]{0cm}{1.1cm}
    \frac{1}{2}\frac{\tan2\t{12}}{|\tan2\t{23}|}(R\cos2\t{12})^{1/2} & 
    \displaystyle\rule[-0.4cm]{0cm}{1.1cm}
    \left(\frac{\sin\t{13}}{\cos2\t{23}}\right)^2 &
    \text{3} &
    &
    \mathbf{1} \\
    \hline
    \hline
    \text{E1} &
    \displaystyle\rule[-0.4cm]{0cm}{1.1cm}
    -\frac{\tan\t{23}}{\cos\delta}\frac{1-\tan\t{12}}{1+\tan\t{12}} &
    2\cot\t{23}\sin\t{13} &
    \text{2, 3} &
    \surd &
    R_{12}(R_{23}) \\
    \hline
  \end{array}\]
  \mycaption{Summary of the possible correlations between $\t{13}$,
  $m_{ee}$ and $\t{23}$, $\t{12}$, $\delta$, 
  $R$, to leading order of the expansion
  in $R$ (with $m_{\text{atm}}\equiv\sqrt{|\dm{32}|}$). The 
  column ``SSC'' gives 
  the number of right-handed neutrinos involved in the see-saw
  realization of each case. An inverse hierarchy is obtained only in
  the case E. Cases A2, B2, E2, are obtained from A1, B1, E1, with 
  the replacements $\tan\t{23} \rightarrow \cot\t{23}$ and 
  $\cos \delta \rightarrow - \cos\delta$.} 
  \label{tab:results}
\end{table}
\ass

\begin{figure}
\begin{flushleft}
\epsfig{file=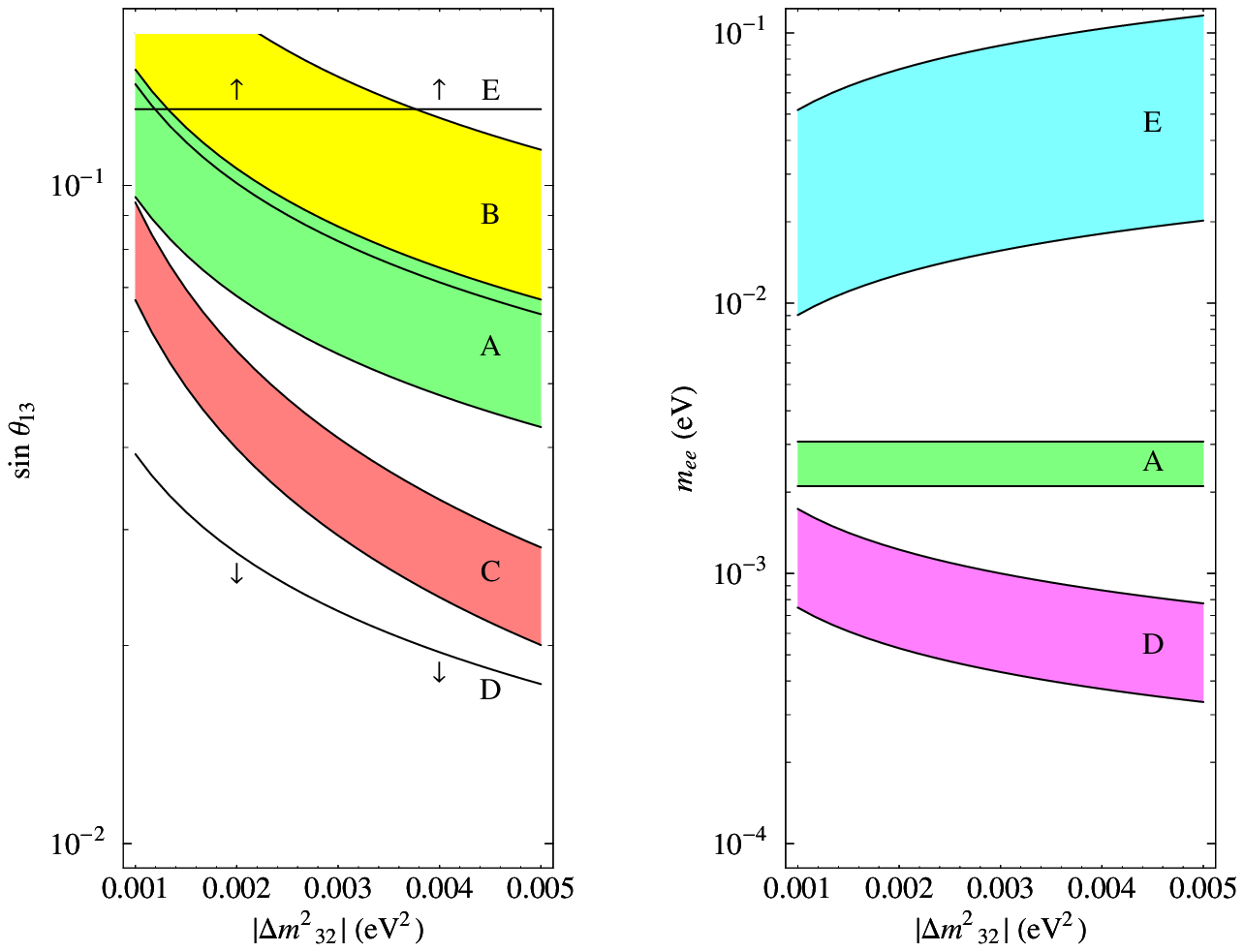,width=0.94\textwidth}
\end{flushleft}
\begin{minipage}[t]{0.475\textwidth}
  \mycaption{Ranges of values for $\sin\t{13}$ at 90\% confidence
    level for the different cases, plotted as a function of $\dm{32}$.
    Cases D, E, which only give a bound on $\sin\t{13}$, are shown 
    with a double arrow. \label{fig:s13}}
\end{minipage}\hspace*{0.05\textwidth}
\begin{minipage}[t]{0.475\textwidth}
  \mycaption{Ranges of values for $m_{ee}$ at 90\% confidence
  level. \label{fig:mee}} 
\end{minipage}
\end{figure}

Our results for the NSSC and for the SSC with 2 and 3 right-handed
neutrinos are summarized in Table~\ref{tab:results} and illustrated in
Fig.~\ref{fig:s13} and Fig.~\ref{fig:mee}. Table~\ref{tab:results} is
meant to be self explanatory. There we show the definite testable
correlations between $\t{13}$, $m_{ee}$ and $\t{23}$, $\t{12}$,
$\delta$ and $R$, that can arise from all possible cases, both in the
NSSC and in the SSC, and are consistent with present data. There are
only four cases (A, B, C, D) that allow to connect $\sin\t{13}$ and
$m_{ee}$ with $\t{23}$, $\t{12}$ and $R$, whereas in case E the
correlation also involves the CP-violating phase $\delta$.  Quite a
few ``natural'' cases have been left out from Table~\ref{tab:results}
because they are not compatible with current data at 90\% confidence
level: some give $\t{12} \geq 45^\circ$, while two predict a too large
value of $\sin\t{13}$, $\sin\t{13} = \sin\t{12} R^{1/4}$ or
$\sin\t{13} = \tan\t{12}/2 (R\cos\t{12})^{1/4}$. One case has been
omitted because it is consistent, within the present uncertainty, with
$\t{13}=0$ (F in the following).

The relations quoted in Table~\ref{tab:results} are obtained through
an expansion in $R$. The higher orders in the expansion are suppressed
by $R^{1/2}$ in all cases except $E$, where the leading corrections to
the relations in Table~\ref{tab:results} are of order $R$.  In all
cases, anyhow, the exact relations can be obtained from
Tables~\ref{tab:SSC2}, \ref{tab:SSC3}, and~\ref{tab:NSSC}, where the
sets of parameters that originate these correlations are shown. They
will be motivated and discussed in Sect.~\ref{sec:explanation}.  All
cases can be obtained in the see-saw context. Table~\ref{tab:results}
specifies the number of right-handed neutrinos involved and whether
the individual cases can also originate from the NSSC. It also gives
the form of the rotation on the charged lepton sector. The form of the
light neutrino mass matrix before the charged lepton rotation is given
in the Appendix. E is the only case that leads to an ``inverted''
spectrum. For cases A1, B1, E1, the independent possibility exists
where $\tan\t{23} \rightarrow \cot\t{23}$, $\cos\delta \rightarrow
-\cos\delta$, denoted in the following by A2, B2, E2, respectively.
Case A is discussed in
Ref.~\cite{Frampton:02b,Raidal:02a}.\footnote{For recent studies of
  textures in neutrino masses, also in connection with leptogenesis,
  see also~\cite{studies}.}

Given the present knowledge of $\t{23}$, $\t{12}$ and $\dm{21}$,
including the recent Kamland result~\cite{kamland:02a}, the ranges of
values for $\sin\t{13}$ are shown in Fig.~\ref{fig:s13} at 90\%
confidence level for the different cases.  It is interesting that all
the ranges for $\sin\t{13}$, except in case D, are above $\simeq0.02$
and some can saturate the present limit.  Long-baseline experiments of
first or second generation should explore a significant portion of
this range while reducing at the same time the uncertainties of the
different predictions at about 10\% level~\cite{Apollonio:02a}.  Note
that, in cases E, although the determination of $\sin\t{13}$ requires
the knowledge of the CP violating phase as well, the allowed range is
still limited, being $\sin\t{13}\gtrsim 0.10$.  Furthermore, the
requirement of not exceeding the present experimental bound on
$\sin\t{13}$ gives a lower bound on $|\cos\delta|$ (and therefore an
upper limit on CP-violation) that we can quantify as
\begin{equation}
  \label{eq:delta}
  |\cos\delta| > 0.8 \quad\text{at 90 \% CL }
\end{equation}
given the present uncertainties. Notice that $\cos\delta < 0$ ($>0$)
in case E1 (E2).  Verifying the prediction for $\sin\t{13}$ in case D
would require the measurement of $\t{23} \neq 45^\circ$; a bound on
$|1-\sin^22\t{23}|$ only sets an upper bound on $\sin\t{13}$, as shown
in Fig.~1.\footnote{The inclusion of higher corrections in $R$ does
  not change this conclusion in any significant way.}

While the prediction for $\sin\t{13}$ are in an experimentally
interesting range, the expectations for the $0\nu2\beta$-decay
effective mass are mostly on the low side, except, as
expected~\cite{Feruglio:02a}, in the only inverted hierarchical case
E. The ranges for each individual cases with non-vanishing $m_{ee}$
are shown in Fig.~\ref{fig:mee}.  The challenge of detecting a non
zero $m_{ee}$, when applicable, is therefore harder than for
$\sin\t{13}$, with a better chance for the only inverted hierarchical
case E.

\section{Table~\ref{tab:results} justified}
\label{sec:explanation}

In this Section we describe how we arrive to select the relatively
small number of cases enumerated in Sect.~\ref{sec:cases}. We build up
the overall picture by commenting on the individual cases, which are
summarized in Tables~\ref{tab:SSC2}, \ref{tab:SSC3} and
\ref{tab:NSSC}.  It is non trivial to make sure that we are not
missing any possibility. This is obtained partly by general
considerations and partly by direct inspection of the individual
cases.

\subsection{Non See-Saw Case, $M_E$ diagonal}


As an illustration, before describing the positive cases, let us show why
the NSSC with diagonal $M_E$ does not lead to any acceptable example, 
according to our rules. The neutrino mass matrix in the flavour 
basis, $\mnf$, coincides with
$\mn$ in \eq{LNSSC}. This justifies considering the matrices $\mnf$ with
2 vanishing entries also in a natural sense. The only form of $\mnf$
which gives the zeroth order pattern characterized by $R=0$ and
$\t{23}$ large without correlations between the different entries, is
(PD = Pseudo-Dirac)~\cite{Barbieri:98d}
\begin{equation}
  \label{eq:PD}
  \mnf^{\text{PD}} = m_0
\begin{pmatrix}
0 & c & s \\
c & 0 & 0 \\
s & 0 & 0
\end{pmatrix},
\end{equation}
where $m_0$ is the overall mass scale and $c$, $s$ stand for the
cosine and the sine of a mixing angle, without loss of generality.
Perturbations to~(\ref{eq:PD}) can lead to a $\mnf$ consistent with
the data with an inverted hierarchical spectrum, except for one
difficulty: the angle $\t{12}$ is too close to $45^\circ$ degrees. To
get inside the range~(\ref{eq:obs}) requires a fine-tuning.  For this
reason we have not considered these cases here. To get solutions in
the NSSC without fine-tuning requires a non-trivial rotation from the
charged lepton sector (see Table~\ref{tab:NSSC}).\footnote{To account
  for the atmospheric neutrino observations, $s/c$ in \eq{PD} should
  be one within about 15\%. This \emph{is} a relation among parameters
  that we assume here and in the following without offering any
  explanation of it. We also assume that it is worth accounting for
  any other correlation among parameters in a natural way, as defined
  above.}

\subsection{See-Saw Case (2 $N$'s), $M_E$ diagonal}

Again because $\t{12}$ is too close to $45^\circ$, one does not get a
natural solution with an inverted spectrum even in the SSC with a
diagonal $M_E$.  On the other hand, it is easy to obtain a normal
hierarchical pattern. A $2 \times 2$ matrix $\mu$ defined in \eq{mu}
with one dominant diagonal element and a non zero determinant may
lead, in fact, to a fully natural description of the
pattern~(\ref{eq:obs})~\cite{Smirnov:93a,King:98a}. Taking $\mu_{22}$
as the dominant element, without loss of generality, one is readily
convinced that the matrix $A$ must have the form
\begin{equation}
  \label{eq:AH}
  A = 
\begin{pmatrix}
0 & s & c \\
c' & s' e^{i\phi} & 0
\end{pmatrix} 
\quad\text{or}\quad
  A = 
\begin{pmatrix}
0 & s & c \\
c' & 0 & s' e^{i\phi}
\end{pmatrix} ,
\end{equation}
where we explicitly included the only phases ineliminable by a
redefinition of the neutrino fields. By counting the number of real
free parameters, which should not exceed 4, the only ambiguity is
where one places the single small but not vanishing entry in $\mu$
other than $\mu_{22}$. There are 2 possibilities: $\mu_{11}\neq 0$,
which leads to cases A in Table~\ref{tab:results}, and
$\mu_{12}=\mu_{21}\neq 0$, which leads to $\sin\t{13}=\sin\t{12}
R^{1/4}$, outside the currently allowed region. Notice that case A
splits in two phenomenologically similar cases A1 and A2 according to
which of the two possibilities in \eq{AH} is chosen.  Clearly, case A2
can be obtained from case A1 by a $\nu_\mu\leftrightarrow\nu_\tau$
exchange in the mixing matrix, which corresponds to
$\t{23}\leftrightarrow\pi/2-\t{23}$ in the predictions.

\asl
\begin{table}[t]
  \centering
  \[\begin{array}{|c||c|c|c|} 
    \hline
    &
    \mu/\mu_0 &
    A &
    U_l \\
    \hline\hline
    \text{A1} & 
    \begin{pmatrix} \epsilon & 0 \\ 0 & 1 \end{pmatrix} &
    \begin{pmatrix} 0 & s & c \\ c' & s'e^{i\phi} & 0 \end{pmatrix} &
    \mathbf{1} \\
    \text{A2} &
    \begin{pmatrix} \epsilon & 0 \\ 0 & 1 \end{pmatrix} &
    \begin{pmatrix} 0 & s & c \\ c' & 0 & s'e^{i\phi} \end{pmatrix} &
    \mathbf{1} \\
    \hline
    \text{E1 (E2)} &
    \begin{pmatrix} 0 & 1 \\ 1 & 0 \end{pmatrix} &
    \begin{pmatrix} 1 & 0 & 0 \\ 0 & c & s \end{pmatrix} &
    \text{$R_{12}$ ($R_{13}$)} \\
    & \multicolumn{2}{c|}{\text{+ 1 small entry}} & \\
    \hline
  \end{array}\]
  \mycaption{Parameters for the see-saw cases with 2 right-handed
  neutrinos of Table~\ref{tab:results}.  $A$ is the Dirac neutrino mass 
matrix with $(A A^\dagger)_{ii}$ normalized to unity, $M_{RL}=m A$, 
$m_{ij}=m_i \delta_{ij}$ and $\mu$ is related to the right handed 
neutrino mass matrix by $\mu=m^{-1} M_R m^{-1}$. $\epsilon$ and $\sigma$ denote
  small entries relative to unity. c, s or c', s' denote the cosine and the
sine of arbitrary angles, $\theta$ and $\theta'$. $U_l$ is 
the rotation of the left handed charged leptons.}
  \label{tab:SSC2}
\end{table}
\ass

\subsection{See-Saw Case (3 $N$'s), $M_E$ diagonal}

In the SSC with 3 $N$'s and a diagonal $M_E$, the atmospheric angle
can originate either from the heavy neutrinos Majorana mass matrix
$M_R$ or from the Dirac mass matrix $M_{RL}$.  In order to give rise
to a large $\t{23}$, the right-handed neutrino mass matrix must
correspond to a $\mu$ in the form
\begin{equation}
  \label{eq:muPD}
  \mu/\mu_0 = 
  \begin{pmatrix}
    a & c & s \\
    c & 0 & 0   \\
    s & 0 & 0
  \end{pmatrix}
  +\text{1 small entry,}
\end{equation}
where $a=\ord{1}$ or $a=0$.  The large $\t{12}$ angle and a
non-vanishing $\dm{21}$ can also be obtained from~(\ref{eq:muPD}) by
adding a small entry in the ``23'' submatrix, provided that $a\neq 0$.
No special structure is demanded to the Dirac matrix, which has to be
diagonal in this case ($A=\mathbf{1}$) since the four available
parameters have been already used in $\mu$.\footnote{The texture in which
  the small entry in~\eq{muPD} is in the 22 or 33 position appears also
  in~\cite{Kageyama:02a}.} On the other hand, if $a=0$, the
solar mixing angle must be provided by a non-diagonal Dirac matrix:
\begin{equation}
  \label{eq:A1}
  A = 
  \begin{pmatrix}
    c' & s'e^{i\phi} & 0 \\
    0 & 1 & 0 \\
    0 & 0 & 1
  \end{pmatrix}
  \quad\text{or}\quad
  \begin{pmatrix}
    c' & 0 & s'e^{i\phi} \\
    0 & 1 & 0 \\
    0 & 0 & 1
  \end{pmatrix}.
\end{equation}
A non-vanishing entry in the ``23'' submatrix of $\mu$ is still
necessary in order to have a non zero determinant\footnote{The small
  determinant limit is in principle as meaningful as the large
  determinant limit leading to the SSC with 2 $N$'s. However, it does
  not give rise to any interesting case.}.

As for the cases in which the $\t{23}$ rotation arises from the Dirac
mass matrix, one needs
\begin{equation}
  \label{eq:mu2}
  \mu/\mu_0 = 
  \begin{pmatrix}
    a & 1 & 0  \\
    1 & 0 & 0  \\
    0 & 0 & \sigma 
  \end{pmatrix}
\end{equation}
Again, the solar angle can originate from~(\ref{eq:mu2}) if $a\neq
0$\,\footnote{Here, as above, we have conventionally chosen a
  labeling for the three right-handed neutrinos that corresponds to a
  given order of the rows in the matrix $A$.} and
\begin{equation}
  \label{eq:A2}
  A = 
  \begin{pmatrix}
    1 & 0 & 0 \\
    0 & 1 & 0 \\
    0 & s & c
  \end{pmatrix}
  \quad\text{or}\quad
  \begin{pmatrix}
    1 & 0 & 0 \\
    0 & 0 & 1 \\
    0 & s & c
  \end{pmatrix}.
\end{equation}
Otherwise, if $a = 0$, the solar angle must also be provided by a
matrix $A$ in the form
\begin{equation}
  \label{eq:A3}
  \begin{pmatrix}
    c' & s'e^{i\phi} & 0 \\
    0 & 1 & 0 \\
    0 & s & c
  \end{pmatrix}
  \;\text{or}\;
  \begin{pmatrix}
    c' & 0 & s'e^{i\phi} \\
    0 & 1 & 0 \\
    0 & s & c
  \end{pmatrix}
  \;\text{or}\;
  \begin{pmatrix}
    c' & s'e^{i\phi} & 0 \\
    0 & 0 & 1 \\
    0 & s & c
  \end{pmatrix}
  \;\text{or}\;
  \begin{pmatrix}
    c' & 0 & s'e^{i\phi} \\
    0 & 0 & 1 \\
    0 & s & c
  \end{pmatrix}\!.
\end{equation}
All the SSCs with 3 $N$'s and a diagonal $M_E$ are summarized in
Table~\ref{tab:SSC3}, when they lead to a case consistent with current
observations. Cases B2 can be obtained from cases B1 by exchanging the
last two columns of $A$ (and possibly a relabeling of $N_1$, $N_2$,
$N_3$). The corresponding predictions are reported in
Table~\ref{tab:results}.

\asl
\begin{table}[t]
  \centering
  \[\begin{array}{|c||c|c|c|} 
    \hline
    &
    \mu/\mu_0 &
    A &
    U_l \\
    \hline\hline
    \text{B1 (B2)} &
    \begin{pmatrix}
      a & c & s \\ c & 0\,(\epsilon e^{i\phi}) & 0 \\ 
      s & 0 & \epsilon e^{i\phi}\, (0)
    \end{pmatrix} &
    \begin{pmatrix}
      1 & 0 & 0 \\ 0 & 1 & 0 \\ 0 & 0 & 1
    \end{pmatrix} &
    \mathbf{1} \\
    \text{} &
    \begin{pmatrix}
      0 & c & s \\ c & 0\,(\sigma) & 0 \\ s & 0 & \sigma \,(0)
    \end{pmatrix} &
    \begin{pmatrix}
      c' & s'e^{i\phi} & 0 \\ 0 & 1 & 0 \\ 0 & 0 & 1
    \end{pmatrix} \!,\;
    \begin{pmatrix}
      c' & 0 & s'e^{i\phi} \\ 0 & 1 & 0 \\ 0 & 0 & 1
    \end{pmatrix} &
    \mathbf{1} \\
    \text{} &
    \begin{pmatrix}
      ae^{i\phi} & 1 & 0 \\ 1 & 0 & 0 \\ 0 & 0 & \sigma
    \end{pmatrix} &
    \begin{pmatrix}
      1 & 0 & 0 \\ 0 & 1 \, (0) & 0 \, (1) \\ 0 & s & c
    \end{pmatrix} &
    \mathbf{1} \\
    \text{} &
    \begin{pmatrix}
      0 & 1 & 0 \\ 1 & 0 & 0 \\ 0 & 0 & \sigma
    \end{pmatrix} &
    \begin{pmatrix}
      c' & s'e^{i\phi} & 0 \\ 0 & 1\,(0) & 0\, (1) \\ 0 & s & c
    \end{pmatrix} \!,\;
    \begin{pmatrix}
      c' & 0 & s'e^{i\phi} \\ 0 & 1 \, (0) & 0 \, (1) \\ 0 & s & c
    \end{pmatrix} &
    \mathbf{1} \\
    \hline
    \text{C} &
    \begin{pmatrix}
      0 & 1 & 0 \\ 1 & 0 & 0 \\ 0 & 0 & \sigma
    \end{pmatrix} &
    \begin{pmatrix}
      c' & 0 & s'e^{i\phi} \\ 0 & 1 & 0 \\ 0 & 0 & 1
    \end{pmatrix} &
    R_{23} \\
    \hline
     \text{D} &
    \begin{pmatrix}
      a & c & s \\ c & 0 & \epsilon e^{i\phi} \\ s & \epsilon
      e^{i\phi} & 0
    \end{pmatrix} &
    \begin{pmatrix}
      1 & 0 & 0 \\ 0 & 1 & 0 \\ 0 & 0 & 1
    \end{pmatrix} &
    \mathbf{1} \\
    \text{} &
    \begin{pmatrix}
      0 & c & s \\ c & 0 & \sigma \\ s & \sigma & 0
    \end{pmatrix} &
    \begin{pmatrix}
      c' & s'e^{i\phi} & 0 \\ 0 & 1 & 0 \\ 0 & 0 & 1
    \end{pmatrix} \!,\;
    \begin{pmatrix}
      c' & 0 & s'e^{i\phi} \\ 0 & 1 & 0 \\ 0 & 0 & 1
    \end{pmatrix} &
    \mathbf{1} \\
    \hline
    \text{E1} &
    \begin{pmatrix}
      0 & \sigma & 0 \\ \sigma & 0 & 0 \\ 0 & 0 & 1
    \end{pmatrix} &
    \begin{pmatrix}
      0 & c & s \\ 1 & 0 & 0 \\ 1 & 0 & 0
    \end{pmatrix}\!,
    \begin{pmatrix}
      0 & c & s \\ 1 & 0 & 0 \\ 0 & 1 & 0
    \end{pmatrix}\!,
    \begin{pmatrix}
      0 & c & s \\ 1 & 0 & 0 \\ 0 & 0 & 1
    \end{pmatrix} &
    R_{12} \\
    \hline
    \text{E2} &
    \multicolumn{2}{c|}{\text{same as E1}} & 
    R_{13} \\
    \hline
\end{array}\]
  \mycaption{Parameters for the see-saw cases with 3 right-handed
  neutrinos of Table~\ref{tab:results}. $\epsilon$ and 
$\sigma$ denote small positive quantities, while $a=\ord{1}$. $\mu$,
$A$ and $U_l$ as in Table~\ref{tab:SSC2}.}
  \label{tab:SSC3}
\end{table}
\ass

\subsection{$M_E$ non diagonal}

With $M_E$ non-diagonal, the mixing matrix $\U = U_l^\dagger U_\nu$
receives a non trivial contribution also from the rotation matrix in
the charged sector $U_l$. Within our hypotheses, none of the 4 general
parameters in $U_l$ can be correlated with the charged lepton masses.
It is on the contrary possible to obtain an arbitrary $U_l$ with the
only limitation that a large rotation in the 23 sector, $R^l_{23}$, if
present at all, should precede all the rotations in the other sectors
$R^l_{12}$ and $R^l_{13}$. Explicitly $U_l=R^l_{23}R^l_{13}R^l_{12}$,
up to phases, is a natural form, with any of the $R^l_{ij}$ possibly
reduced to $\mathbf{1}$. This follows from the hierarchy of the
charged lepton masses, ordered in the usual way, $(1,2,3) =
(e,\mu,\tau)$. Note that $R^l_{12}$ must be close to the identity
since otherwise, when commuted with $R^l_{23}$ to obtain the
conventional order in $\U$, a large $R_{13}$ rotation would also be
generated.

\subsubsection{$U_l=R^l_{23}$}

$U_l=R^l_{23}$ (up to phases) must be considered in association with
the NSSC and SSC cases for $\mn$. By inspection one shows that in both
cases it is only in association with a hierarchical pattern of
neutrino masses that interesting cases can be produced, if we
disregard the further generation of cases with inverted spectrum and
the $\t{12}\simeq 45^\circ$ problem already encountered. The only
predictive and viable form of $\mn$ is
\begin{equation}
  \label{eq:NSSCR23}
  \mn = \begin{pmatrix}
0 & \sigma &0  \\
\sigma & 0 & \epsilon \\
0 & \epsilon & 1
\end{pmatrix}
\end{equation}
($\sigma$, $\epsilon$ are small corrections), which leads to case C in
Table~\ref{tab:results} and \ref{tab:NSSC}.

\asl
\begin{table}[t]
  \centering
  \[\begin{array}{|c||c|c|} 
    \hline
    &
    \mn/m_0 &
    U_l \\
    \hline\hline
    \text{C} &
    \begin{pmatrix} 0 & \sigma & 0 \\ \sigma & 0 & \epsilon \\
    0 & \epsilon & 1 \end{pmatrix} &
  R_{23} \\
  \hline
    \text{E1 (E2)} &
    \begin{array}{c}
      \begin{pmatrix} 0 & c & s \\ c & 0 & 0 \\
        s & 0 & 0 \end{pmatrix} \\
      \text{+1 small entry}
      \end{array} &
  \text{$R_{12}$ ($R_{13}$)} \\
    \hline
  \end{array}\]
  \mycaption{Parameters for the non see-saw case of
  Table~\ref{tab:results}. $\epsilon$ and $\sigma$ denote small 
  entries relative to unity. $M_N$ is the left handed neutrino mass matrix.
  $U_l$ as in Table~\ref{tab:SSC2}.}
  \label{tab:NSSC}
\end{table}
\ass

In the SSC with 2 $N$'s, a $R^l_{23}$ contribution from the charged
lepton sector opens the possibility
\begin{equation}
  \label{eq:SSCU232}
  \mu = \mu_0 \begin{pmatrix}
\sigma & 0 \\
0 & 1 
\end{pmatrix}, \qquad
A = 
\begin{pmatrix}
0 & 0 & 1 \\
c' & s' & 0
\end{pmatrix}.
\end{equation}
This gives a variation of case A which predicts $\sin\t{13}=0$ and
$m_{ee}$ as in the other cases A, probably too small to be detected
and, as such, not included in the summary of Table~\ref{tab:results}.

With 3 $N$'s, the only viable possibility is
\begin{equation}
  \label{eq:muA}
  \mu/\mu_0 = 
  \begin{pmatrix}
    0 & 1 & 0 \\
    1 & 0 & 0 \\
    0 & 0 & \sigma
  \end{pmatrix} \!, \quad A=
  \begin{pmatrix}
    c' & 0 & s'e^{i\phi} \\ 0 & 1 & 0 \\ 0 & 0 & 1
  \end{pmatrix}
\end{equation}
leading to case C again. The case
\begin{equation}
  \label{eq:last}
  \mu/\mu_0 = 
  \begin{pmatrix}
    0 & \sigma & 0 \\
    \sigma & 0 & 0 \\
    0 & 0 & 1
  \end{pmatrix}\!, \quad A=
  \begin{pmatrix}
    0 & 0 & 1 \\
    0 & 0 & 1 \\
    s' & c' & 0
  \end{pmatrix}
\end{equation}
gives the same predictions as~(\ref{eq:SSCU232}).

Since a $\nu_\mu\leftrightarrow\nu_\tau$ exchange can be reabsorbed in
a redefinition of the $\t{23}$ mixing angle, the cases discussed in
this subsection do not split in subcases. The experimental observation
that $\tan^2\t{12}$ not only deviates from 1 but is lower than 1
serves to cut out a few otherwise acceptable cases.

\subsubsection{$R^l_{12}$ or $R^l_{13}\neq 0$}
\label{sec:1213}

As already mentioned, the rotations in $U_l$ other than $R_{23}^l$, if
present at all, must be small. On the other hand, they introduce new
parameters. Therefore, the only way in which they can give rise to a
new realistic and predictive case is when they act in association with
a $\mn$ that depends at most on 3 effective parameters and is close
enough to the fully realistic situation. On this basis, one is readily
convinced that an interesting case is represented by any $\mn$,
frequently encountered, which gives an inverted spectrum and
$\t{12}^\nu\simeq 45^\circ$.  The role of the charged lepton
contribution $R^l_{12}$ or $R^l_{13}$ is essential in providing the
required correction to $\t{12}=45^\circ$ in a natural way.

In this way one is led to consider mixing matrices of the form
\begin{equation}
  \label{eq:U12}
  \U = R^l_{12}(\alpha)R_{23}
\begin{pmatrix}
e^{i\phi} & & \\
& 1 & \\
& & 1
\end{pmatrix}
R_{12}(\t{12}^\nu)
\end{equation}
where $\alpha$ is a free small mixing parameter. Note that the case of
a $R^l_{13}$ contribution from the lepton sector can be obtained from
\eq{U12} with a $\nu_\mu\leftrightarrow\nu_\tau$ exchange (and a
redefinition of $R_{23}$), which corresponds to
$\t{23}\leftrightarrow\pi/2-\t{23}$ and $\delta\leftrightarrow\delta +
\pi$ in the predictions. To obtain $\t{13}$, it is necessary to bring
$\U$ to the standard form, i.e.\ to commute the $R^l$ rotations,
either $R^l_{12}$ or $R^l_{13}$, with $R_{23}$. To first order in
$\alpha$, it is
\begin{equation}
  \label{eq:comm}
  R^l_{12}(\alpha)R_{23}
\begin{pmatrix}
e^{i\phi}\hspace*{-7.5pt} & & \\
& 1 & \\
& & 1
\end{pmatrix}
R_{12}(\t{12}^\nu) = 
R_{23} R_{13}(\alpha s_{23})\begin{pmatrix}
e^{i\phi}\hspace*{-7.5pt} & & \\
& 1 & \\
& & 1
\end{pmatrix}
R_{12}(\t{12}^\nu+\alpha c_{23}\cos\phi)
\end{equation}
which leads to case E1 in Table~\ref{tab:results} and \ref{tab:SSC3}
(where the relations are exact in $\alpha$). The corresponding case E2
is obtained for $U_l=R^l_{13}$.

The explicit realizations of the mechanism are obtained in the NSSC
case by adding a single correction to $\mn$ in \eq{PD}. In the SSC
case with 2 $N$'s, they correspond to adding a single correction to
$\mu/\mu_0$ or $A$ in
\begin{equation}
  \label{eq:SSCU121}
  \mu = \mu_0 \begin{pmatrix}
0 & 1 \\
1 & 0 
\end{pmatrix}\!,\quad
A = 
\begin{pmatrix}
1 & 0 & 0 \\
0 & c & s
\end{pmatrix}\!, 
\end{equation}
while in the SCC with 3 $N$'s they correspond to 
\begin{equation}
  \label{eq:muA3}
  \mu/\mu_0 = 
  \begin{pmatrix}
    0 & \sigma & 0 \\
    \sigma & 0 & 0 \\
    0 & 0 & 1
  \end{pmatrix} \,\text{and}\; A =
  \begin{pmatrix}
    0 & c & s \\ 1 & 0 & 0 \\ 1 & 0 & 0
  \end{pmatrix} \!,\; 
  \begin{pmatrix}
    0 & c & s \\ 1 & 0 & 0 \\ 0 & 1 & 0
  \end{pmatrix} \!\text{, or}\; 
  \begin{pmatrix}
    0 & c & s \\ 1 & 0 & 0 \\ 0 & 0 & 1
  \end{pmatrix}\!.
\end{equation}
Starting from similar configurations, $U_l=R^l_{12}R^l_{23}$ and
$U_l=R^l_{13}R^l_{23}$ lead to the same cases as $U_l=R^l_{12}$ and
$U_l=R^l_{13}$.

The leptonic rotations $R^l_{12}$ and $R^l_{13}$ can also play a role
in hierarchical cases in which the neutrino contribution to $\U$ gives
$\t{12}>45^\circ$. The charged lepton contribution is then used to
bring $\t{12}$ back to the allowed range below $45^\circ$. This
possibility is only realized in the SSC with 3 $N$'s for
\begin{equation}
  \label{eq:muA4}
  \mu/\mu_0 = 
  \begin{pmatrix}
    0 & 1 & 0 \\
    1 & 0 & 0 \\
    0 & 0 & \sigma
  \end{pmatrix}\,\text{and}\; A =
  \begin{pmatrix}
    1 & 0 & 0 \\ 0 & 1 & 0 \\ 0 & s & c
  \end{pmatrix} \!,\; 
  \begin{pmatrix}
    1 & 0 & 0 \\ 0 & 0 & 1 \\ 0 & s & c
  \end{pmatrix}\!,
\end{equation}
giving respectively cases F1a, F1b for $U_l = R_{12}$ and cases F2a,
F2b for $U_l = R_{13}$; or adding a single correction in the 2-3
submatrix of $\mu/\mu_0$ in
\begin{equation}
  \label{eq:muA5}
  \mu/\mu_0 = 
  \begin{pmatrix}
    0 & c & s \\
    c & 0 & 0 \\
    s & 0 & 0
  \end{pmatrix} \!,\; A =
  \begin{pmatrix}
    1 & 0 & 0 \\ 0 & 1 & 0 \\ 0 & 0 & 1
  \end{pmatrix}\!,
\end{equation}
giving cases F1a, F1b, F1c for $U_l = R_{12}$ and cases F2a, F2b or
F2c for $U_l = R_{13}$.  As anticipated in Section 2, cases F do not
appear in Table 1 because their predictions for $\sin\t{13}$ are
either inconsistent with present data (Fa, Fc) or not conclusive,
since it is at present compatible with zero (Fb). To get to this
conclusion it is necessary to take into account not only the leading
order prediction for $\sin\t{13}$, which is common to the three cases
F1a, F1b, F1c,
\begin{equation}
-\cos\delta \sin\t{13}^{(0)} = \tan\t{23}\,\frac{1-\tan
  \theta_{12}}{1+\tan \theta_{12}} 
\end{equation}
but to include as well higher order corrections in $R$:
\begin{equation}
\begin{split}
-\cos \delta \sin \t{13}  &= \tan \theta_{23} \,
\frac{t-\tan\t{12}}{1+t \tan\t{12}} +
q \left(\frac{R}{2}  \right)^{1/3}   \\
t&=1+\frac{p}{2}\left(\frac{R}{2}\right)^{1/3}, \qquad \text{with} \\
p&=(\tan\t{23})^{4/3}\text{, $(\tan \t{23})^{-4/3}$, $1$} \\
q&=(\tan \t{23})^{1/3}\text{, $-(\tan\t{23})^{-1/3}$, $(\tan2\t{23})^{-1}$}
\end{split}
\end{equation}
in cases F1a, F1b, F1c respectively.

\section{Conclusions}

The economy in the number of basic parameters could be at the origin
of some correlations between the physical observables in the neutrino
mass matrix. At the present state of knowledge, the variety of the
possibilities for the basic parameters themselves is large: $\mn$,
$M_E$, $M_R$, $M_{RL}$ are the matrices that might be involved.
Finding the minimal cases that describe the present pattern of the
data in a natural way could be a first step in the direction of
discriminating the relevant $\L$. This we have done with the results
summarized in Tables~\ref{tab:results}--\ref{tab:NSSC} and illustrated
in \Fig{s13} and \Fig{mee}.  It is remarkable that the number of
possible correlations between the physical observables is limited
(Table~\ref{tab:results}), with a relatively larger number of
possibilities for the basic parameters
(Tables~\ref{tab:SSC2}--\ref{tab:NSSC}). The relatively best chance
for selecting experimentally one out of the few relevant cases is
offered by $\sin\t{13}$. Its predictions, in Table~\ref{tab:results},
should have a 10\% uncertainty with the improved determination of the
other parameters foreseen in long-baseline neutrino
experiments~\cite{Apollonio:02a}. Combining this with independent
studies of leptogenesis or of lepton flavour violating effects could
lead to the emergence of an overall coherent picture.

We have insisted on ``naturalness'' both in solving the ``large
$\t{23}$-small $R$'' problem for the normal hierarchy case and in
obtaining a significant deviation of $\t{12}$ from $45^\circ$, with
small $R$, in the inverted hierarchy case. Both features might be due
to an accidental tuning of parameters. Nevertheless, explaining these
features in a natural way offers a possible interesting guidance for
model building. Note, in this respect, the use in cases E1, E2 of a
relatively small charged lepton rotation in the 12 or 13 sectors to
solve the second fine tuning problem mentioned above. This is
analogous to the use of a large charged lepton rotation in the 23
sector to account for the ``large $\t{23}$-small $R$'' problem in a
natural way~\cite{Albright:98a}.

\section*{Acknowledgments}

This work has been partially supported by MIUR and by the EU under TMR
contract HPRN--CT--2000--00148. Part of the work of A.R. was done
while at the Scuola Normale Superiore. We thank A. Strumia for useful
comments.

\newpage
\begin{appendix}
\section{General forms of the $M_N$ matrices for the various cases}

\asl
\begin{table}[h]
 \centering
  \[\begin{array}{|c||c|c|} 
    \hline
    &
    M_N & U_l\\
    \hline\hline
     \text{A1} &
    \begin{pmatrix}
      \rho^2 & \rho \gamma & 0 \\ 
      \rho \gamma & s^2+\gamma^2 & s c \\ 0 & s c & c^2
    \end{pmatrix}  &
     \mathbf{1} \\
    \hline
    \text{A2} &
    \begin{pmatrix}
      \rho^2 & 0 & \rho \gamma  \\ 
      0  & s^2 & s c \\ \rho \gamma  & s c & c^2+ \gamma^2
    \end{pmatrix}  &
     \mathbf{1} \\
    \hline
    \text{B1} &
    \begin{pmatrix}
      0 & \rho & 0 \\ \rho & s^2+\gamma & s c \\ 0 & s c & c^2
    \end{pmatrix} & 
     \mathbf{1}  \\
    \hline
    \text{B2} &
    \begin{pmatrix}
      0 & 0 & \rho \\ 0 & s^2 & s c \\ \rho & s c & c^2+\gamma
    \end{pmatrix} &
     \mathbf{1}\\
    \hline
    \text{C} &
    \begin{pmatrix}
      0 & \rho & 0 \\ \rho & 0 & \gamma \\ 0 & \gamma & 1
    \end{pmatrix} &
    R_{23}\\
    \hline
    \text{D} &
    \begin{pmatrix}
      \rho^2 & s \rho & - c \rho \\ s \rho & s^2 & s c +\gamma \\ 
       -c \rho & s c + \gamma & c^2
    \end{pmatrix}  &
     \mathbf{1} \\
    \hline
    \text{E1} &
    \begin{pmatrix}
      \rho & c & s \\ c & 0 & 0 \\ s & 0 & 0
    \end{pmatrix}\!,
    \begin{pmatrix}
      0 & c & s \\ c & \rho & 0 \\ s & 0 & 0
    \end{pmatrix}\!,
    \begin{pmatrix}
      0 & c & s \\ c & 0 & 0 \\ s & 0 & \rho
    \end{pmatrix} &
    R_{12}  \\
    \text{} &
    \begin{pmatrix}
      0 & c & s \\ c & 2 \rho c & \rho s \\ 
      s & \rho s & 0
    \end{pmatrix}\!,
    \begin{pmatrix}
      0 & c & s \\ c & 0 & \rho c \\ 
      s & \rho c & 2 s \rho
    \end{pmatrix}\!,
    \begin{pmatrix}
      0 & c & s \\ c & \rho c^2 & \rho s c \\ 
      s &  \rho s c & \rho s^2
    \end{pmatrix} &
    \\
    \hline
    \text{E2} &
    {\text{same as E1}} &
    R_{13}  \\
    \hline
\end{array}\]
  \mycaption{General forms of the $M_N$ matrix (up to an overall
  factor) for the see-saw 
  cases. For $U_l= \mathbf{1}$, this matrix coincides with the neutrino mass 
  matrix $M_\nu$ in the flavor basis. For $U_l \neq \mathbf{1}$, both matrices
  are related by \eq{rotations}, i.e. $M_\nu=U_l^T M_N U_l$. In these
  exact forms, up to a
  sign, the parameters $s$ and $c$ are approximately the same as those in
  Tables~\ref{tab:SSC2}, \ref{tab:SSC3}. $\rho$ can be chosen
  positive, $\gamma$ is in 
  general complex and $|\gamma|\sim\rho\ll 1$.}
  \label{tab:SSC6}
\end{table}
\ass

\end{appendix}


\end{document}